\documentclass[
 prc,
 preprint,
 showpacs,amsmath,amssymb,aps,floatfix,
 superscriptaddress]{revtex4}
 \usepackage{graphicx} % Include figure files
 \usepackage{dcolumn}  % Align table columns on decimal point
 \usepackage{bm}       % bold math
 \usepackage{amssymb}
 \usepackage{amsmath}

\begin{document}
\title{
 Functional medium-dependence of the nonrelativistic optical model potential
}
\author{H. F. Arellano}
\email{arellano@dfi.uchile.cl}
\homepage{http://www.omp-online.cl}
\affiliation{
Department of Physics - FCFM, University of Chile \\
Av. Blanco Encalada 2008, Santiago, Chile} 

\author{Eric Bauge}
\email{eric.bauge@cea.fr}
\affiliation{
Commisariat \`a l'Energie Atomique, 
D\'epartement de Physique Th\'eorique et Appliqu\'ee, 
Service de Physique Nucl\'eaire, 
Boite Postale 12, F-91680 Bruy\`eres-le-Ch\^atel, France}
\date{\today}

%%%%%%%%%%%%%%%%%%%%%%%%%%%%%%%%%%%%%%%%%%%%%%%%%%%%%%%%%%%%%%%%%%%%%
\begin{abstract}
By examining the structure in momentum and coordinate space
of a two-body interaction spherically symmetric in its local coordinate,
we demonstrate that it can be disentangled into two distinctive
contributions.
One of them is a medium-independent and momentum-conserving term,
whereas the other is functionally --and exclusively-- proportional
to the radial derivative of the reduced matrix element.
As example, this exact result was applied to the unabridged optical
potential in momentum space, leading to an explicit separation between
the medium-free and medium-dependent contributions.
The latter does not depend on the strength of the reduced effective
interaction but only on its variations with respect to the density.
The modulation of radial derivatives of the density enhances the effect
in the surface and suppresses it in the saturated volume.
The generality of this result may prove to be useful for the study
of surface-sensitive phenomena.
\end{abstract}

\maketitle

%%%%%%%%%%%%%%%%%%%%%%%%%%%%%%%%%%%%%%%%%%%%%%%%%%%%%%%%%%%%%%%%%%%%%
\section{Introduction}
During the past two decades, several developments in theoretical 
nuclear research have allowed significant improvements in the 
microscopic description of nuclear collisions \cite{Ray92,Amo00}.
Such is the case of nucleon scattering, where continuous efforts 
have led to detailed microscopic realizations of optical model
potentials, thus providing the most complete current first-order 
nonrelativistic description of nuclear collisions off nuclei
over a wide energy range and various targets.
These calculations emphasize a detailed treatment of the NN effective
interaction, particularly its density dependence as well as its
energy and momentum dependence.
Quite generally, however, all realizations of the optical potential 
become a single expression in the form of a convolution of 
medium-dependent effective interactions folded with the target 
ground-state mixed density.
In this article we demonstrate that the optical potential can be
expressed as the sum of two very distinctive terms, one of them 
depends exclusively on the free-space $t$ matrix and the other 
as a gradient of the reduced $g$ matrix.
This result implies that intrinsic medium effects in optical 
potentials become enhanced in the nuclear surface and suppressed 
in the saturated volume. 

Since the early realizations of microscopic optical potentials
\cite{Jeu76,Bri78,Ger84}, the role of nuclear medium effects has been 
a major issue in the study of the dynamics involved in nucleon-nucleus
collisions. 
In these studies density-dependent nucleon-nucleon local 
(\emph{NN}) effective interactions have been developed to represent 
the force between nucleons in the nuclear medium.
The use of these local forces, with suitable local density 
prescriptions, have led to folding optical potentials in coordinate
space which provide reasonable descriptions of \emph{NA} scattering 
data at energies between a few tens of MeV up to near 400 MeV.
Recent developments \cite{Amo00,Ray99}, within the same philosophy, 
have succeeded in including non localities in the optical potential
stemming from the exact inclusion of the exchange term. 
In this case the full mixed density from nuclear shell models 
are used and provide reasonable account of the existing \emph{NA} 
scattering data.

A slightly different strategy has been followed by Arellano, Brieva 
and Love (ABL), with the realization of folding optical potentials in 
momentum space \cite{Are95,Are02}. 
In their approach genuine nonlocal $g$ matrices, based on the
Brueckner-Bethe-Goldstone infinite nuclear matter model, are 
folded to the ground-state local density of the target. 
As a result, nonlocal potentials are obtained with varying degree of 
success in describing the low and intermediate energy 
data. % \cite{Are07a,Are07b}.

The inclusion of medium effects has also been addressed within the 
spectator expansion \cite{Chi95}, where the coupling between struck 
nucleons and target spectators is taken into account. 
This approach is an extension of the Watson, and Kerman, McManus 
and Thaler theories, with focus is on the many-body propagator 
involved in the (A+1)-body problem.

From a more general prospective, various formal expressions of the
optical potential can be found in the literature
\cite{Wat53,Fes58,Ker59,Fet65,Vil67}. 
Although they may differ in the way they establish contact with the bare 
\emph{NN} potential, they all become a folding expression between the 
target ground state and a generalized two-body effective interaction.
In this article we analyze this general expression and demonstrate
that, regardless of the model of utilized to represent the effective 
interaction, the intrinsic medium effects become manifest in the
nuclear surface. 
The implications of this result are examined in the framework of
an infinite nuclear matter model for the \emph{NN} effective interaction.

This article is organized as follows. In Section II we outline
the general framework, discuss the structure of two-body operators
and introduce the `asymptotic separation' for spherically symmetric
systems. The result is then applied to the unabridged optical potential
in momentum space.
In Section III we make use of an infinite nuclear matter model for the
effective \emph{NN} interaction and examine its implications in the
optical potential. Furthermore, we analyze its consistency with the
ABL approach, and assess the medium sensitivity of selected matrix 
elements at various energies.
In Section IV we present a summary and the main conclusions of this work.
Additionally, we have added three appendices where we include some
intermediate steps.

%%%%%%%%%%%%%%%%%%%%%%%%%%%%%%%%%%%%%%%%%%%%%%%%%%%%%%%%%%%%%%%%%%%%%
\section{The optical potential}
A general representation of the optical model potential for 
collisions of a hadronic probe with kinetic energy $E$ off a composite
target is given by the expression
\begin{equation}
\label{omp}
U({\bm k}',{\bm k}) = \int d{\bm p}'\;d{\bm p}\; 
\langle {\bm k}' {\bm p}' \mid \hat T\mid {\bm k}\;{\bm p}\;\rangle
\; \hat\rho({\bm p}',{\bm p})\;,
\end{equation}
where $\hat T$ represents a two-body effective interaction containing,
in general, information about the discrete spectrum of the many-body system.
The one-body mixed density $\hat\rho({\bm p}',{\bm p})$ represents 
the ground-state structure of the target.
Thus, a fully consistent evaluation of the optical potential by means of
the full $\hat T$ matrix would require the solution of the $(A+1)$-body system, 
a formidable task.
This difficulty is circumvented by treating separately the 
ground state and the two-body effective interaction. 
This separation becomes suitable at intermediate and high energies,
where the discrete spectrum of the many-body Green's function is distant 
from the projectile energy in the continuum.
Then, the target ground-state may be described resorting to alternative 
framework such as Hartree-Fock-Bogoliubov.
The effective interaction, in turn, can be modeled using the
Brueckner-Bethe-Goldstone approach.

\subsection{Two-body effective interaction}
\begin{figure}[ht]
\includegraphics[scale=0.4,angle=00] {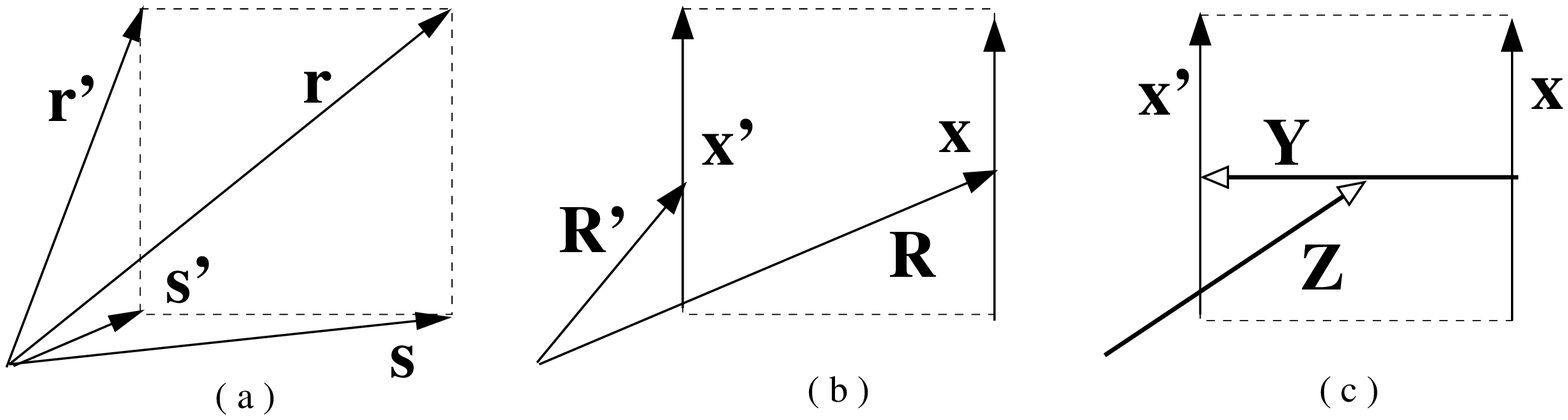}
\medskip
\caption{{\protect\small
\label{fourv}
         Schematic representation of the vector coordinates in a two-body
         operator.}
        }
\end{figure}

\begin{figure}[ht]
\includegraphics[scale=0.45,angle=00] {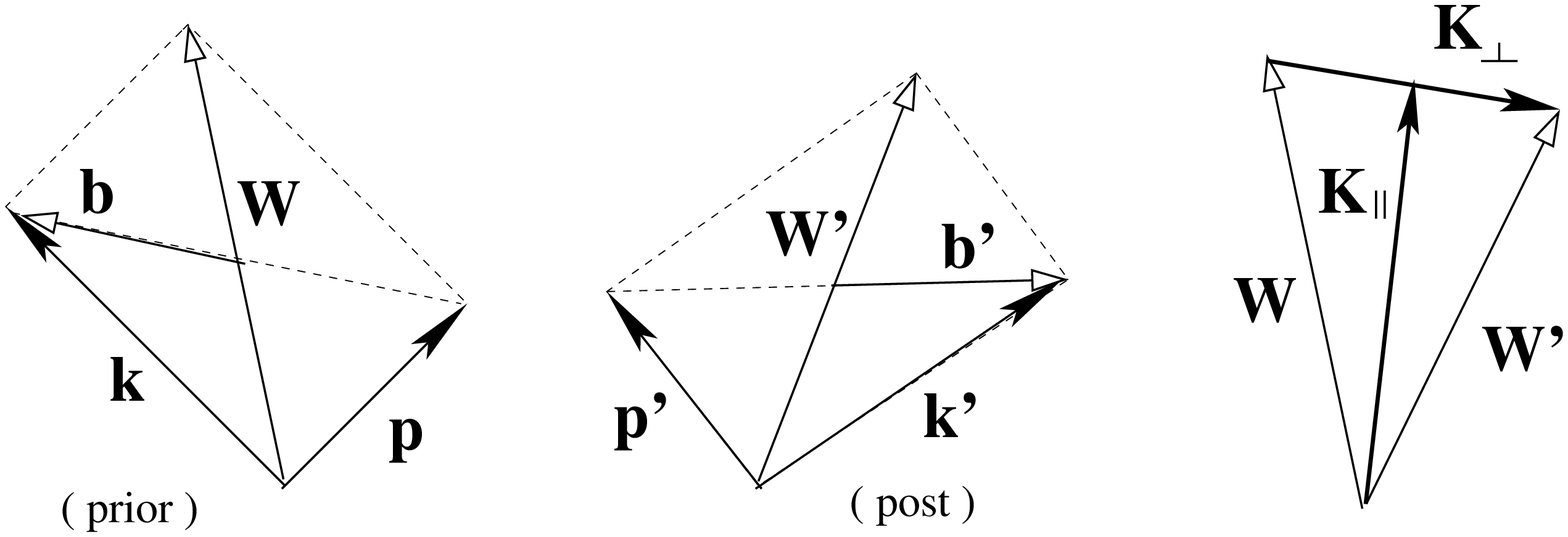}
\medskip
\caption{{\protect\small
\label{fourm}
         Schematic representation of the momenta in a two-body
         operator.}
        }
\end{figure}

Let us first focus our discussion on the two-body effective interaction
and examine its structure.
Quite generally, the representation of the two-body operator $\hat T$ 
in either momentum or coordinate space requires the specification 
of four vectors.
We denote the coordinate representation of $\hat T$ in the form
\[
\langle {\bm r}' {\bm s}'\mid \hat T\mid{\bm r}\; {\bm s}\rangle =
T({\bm r}' {\bm s}';{\bm r}\; {\bm s})\;,
\]
where the `prior' coordinates of each particle are ${\bm r}$ 
and ${\bm s}$, respectively. 
Similarly, ${\bm r}'$ and ${\bm s}'$ refer to the `post' coordinates 
of the same particles, as shown in Fig. (\ref{fourv}a).
An alternative set of coordinates is summarized by the transformation
\begin{equation}
\label{rset1}
\begin{array}{cc}
{\bm R}'=({\bm r}'+{\bm s}')/2 \;;\qquad & {\bm x}'=\bm r'- \bm s' \;;\\ \\
{\bm R} =({\bm r} +{\bm s} )/2 \;;\qquad & {\bm x} =\bm r - \bm s  \;;
\end{array}
\end{equation}
where ${\bm x}$ represents the prior relative coordinate of the pair and
${\bm R}$ their respective center of mass, 
as illustrated in Fig. (\ref{fourv}b).
With this transformation we express the equivalence
$
T({\bm r}'{\bm s}';{\bm r}{\bm s})=T_{\bm R'\bm R}({\bm x}',{\bm x})\;.
$
Thus, following the procedure outlined in Appendix \ref{TwoBody} 
we express the momentum space elements 
$\tilde T\equiv\langle{\bm k}'{\bm p}'\mid\hat T\mid{\bm k}\;{\bm p}\rangle$,
in the form
\begin{equation}
\label{wigner}
\tilde T=
\int\;
\frac{d{\bm Z}}{(2\pi)^3}
\;e^{i{\bm Z}\cdot({\bm W}'-\bm W)}\;
g_{\bm Z}[\textstyle{\frac12}(\bm W'+\bm W); {\bm b}',{\bm b}]\;,
\end{equation}
where $g_{\bm Z}$ represents the reduced interaction at the local
coordinate ${\bm Z}$.
Here we have denoted
\[
{\bm W} = {\bm k} +{\bm p}\;, \qquad
{\bm b} = \textstyle{\frac{1}{2}} ({\bm k} -{\bm p})\;, 
\]
the prior total and relative momenta, respectively.
The same construction applies to the post momenta, where prime marks 
are used.
The relationship between these momenta is illustrated in Fig. (\ref{fourm}).
Eq. (\ref{wigner}) for $\tilde T$ also expresses the role of vector ${\bm Z}$,
\[
\bm Z=\textstyle{\frac{1}{4}}(\bm r'+\bm r+\bm s'+\bm s)\;. \nonumber
\]
the center of gravity of the four coordinates of the two particles.
We name this the \emph{local coordinate}, the \emph{locus} where the reduced 
interaction is evaluated. 
Notice that this coordinate is invariant under the permutation of 
coordinates ${\bm r}\rightleftharpoons {\bm s}$.

The above representation of the $\hat T$ matrix displays
very clearly its dependence in terms of the total 
($\bm W$, $\bm W'$) and relative ($\bm b$, $\bm b'$) momenta.
Additionally, the Wigner transform in the ${\bm R}$ and
${\bm R}'$ coordinates restricts further the structure of $T$, 
suggesting the definitions
\begin{subequations}
\begin{eqnarray}
{\bm W}_\perp &\equiv&
{\bm W}'-{\bm W}=(\bm p'-\bm p)-(\bm k-\bm k')\;; \\
{\bm W}_{||}  &\equiv&
\frac{{\bm W}'+{\bm W}}{2}=
\textstyle{\frac12}(\bm k'+\bm k+\bm p'+\bm p)\;. 
\end{eqnarray}
\end{subequations}
The vector ${\bm W}_\perp$ represents the total momentum gained by the
pair upon interaction (${\bm W}'={\bm W}+{\bm W}_\perp$), whereas 
${\bm W}_{||}$ is the average of the prior and post total momenta.
These momenta become perpendicular only if ${\bm W}'^2={\bm W}^2$.
%%%%%%%%%%%%%%%%%%%%%%%%%%%%%%%%%%%%%%%%%%%%%%%%%%%%%%%%%%%%%%%%%%%%%%%
\subsection{Asymptotic separation of $\tilde T$}
\begin{figure}[ht]
\includegraphics[scale=0.30,angle=00] {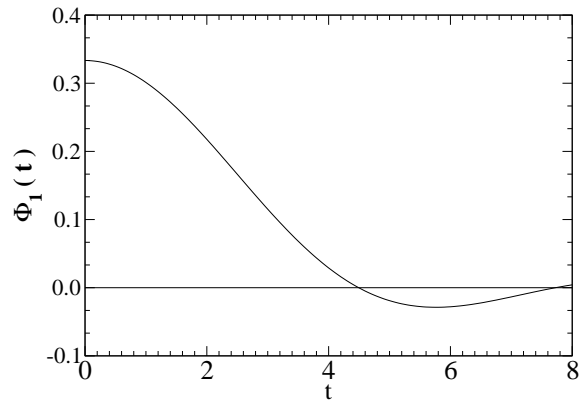}
\bigskip
\caption{{\protect\small
\label{profile}
         The profile function $\Phi_1(t)$ as function of $t$. }
        }
\end{figure}

We now examine the structure of $\tilde T$ in the context of a
finite nucleus with spherical symmetry. 
By that we understand that $g_Z$ depends only on the 
magnitude of the local coordinate, $|\bm Z|=Z$.
Additionally, let us assume that as $Z\to\infty$, $g_Z$ tends
to its free-space form $g_\infty$.
If we omit the three vector arguments of $g_Z$ and decompose it as
$g_Z=(g_Z-g_\infty)+g_\infty$, then
\[
\tilde T=\delta({\bm W}_\perp)g_\infty+
\frac{1}{(2\pi)^3}
\int
d{\bm Z}\;
e^{i{\bm Z}\cdot{\bm W}_\perp}\;
(g_Z-g_\infty)\;.
\]
Carrying out the solid angle integration, 
the integrand of the remaining radial integral is simply 
$4\pi Z^2j_0(ZW_\perp)(g_Z-g_\infty)$, 
which integrated by parts yields the asymptotic separation
%%%%%%%%%%%%%%%%%%%%%%%%%%%%%%%%%%%%%%%%%%%%%%%%%%%%%%%%%%%%%%%%%%%%%
\begin{equation}
\label{TheResult}
\tilde T=\delta({\bm W}_\perp)g_\infty
{\;-\;} %% Sign!
\frac{1}{2\pi^2}
\int_{0}^{\infty}Z^3\;dZ\;
%\left[
%\frac{j_1(Z\,W_\perp)}{Z\,W_\perp}
%\right]
\Phi_1(Z\,W_\perp)\;
\frac{\partial g_Z}{\partial Z}\;.
\end{equation}
Here $\delta$ denotes the Dirac delta function and
$\Phi_1$ represents the profile function defined by
\[
\Phi_1(t) = j_1(t)/t\;,
\]
with $j_1$ the spherical Bessel function of order 1.
In Fig. (\ref{profile}) we plot $\Phi_1$, where we observe that
its peak value (1/3) occurs at the origin, and that it is mainly 
contained within the region $t\lesssim 4$.

What is interesting about Eq. (\ref{TheResult}) is that it separates 
unambiguously the free-space contribution of the $\hat T$ matrix from
its medium-dependent counterpart. 
On the one hand, the medium dependence enters solely as the gradient 
of the reduced element while the total momentum is not conserved.
On the other hand, the medium-independent contribution does conserve
momentum, as dictated by $\delta({\bm W}_\perp)$.
This contrast is physically consistent with our notion about 
non-translational invariant systems. By introducing a $Z$-dependent
reduced interaction, the two-body $\hat T$ matrix does not expresses
conservation of the total momentum.
The conservation becomes manifest only if $\partial g_Z/\partial Z=0$,
as in the cases of interacting nucleons in infinite nuclear matter or 
free space.
More interestingly, the result displayed by Eq. (\ref{TheResult}) is 
sufficiently general to allow us to model the medium dependence in 
a finite nucleus and justifies the incorporation of medium effects
in distorted wave Born approximations (DWBA).

The application of the asymptotic separation in Eq. (\ref{omp}) for $U$ 
yields some undisclosed features.
Let us first change variables from ${\bm p},{\bm p}'$ to ${\bm P},{\bm Q}$,
\[
{\bm P}=({\bm p}'+{\bm p})/2\;; \qquad
{\bm Q}={\bm p}'- {\bm p}\;;
\]
so that $d{\bm p}'d{\bm p}=d{\bm Q}\,d{\bm P}$. 
These two vectors represent the mean and transferred struck-nucleon 
momenta, and the integration on them accounts for the Fermi motion
of the target nucleons. 
Analogously, let us denote
\[
{\bm K}=({\bm k}+{\bm k}')/2\;;\qquad
{\bm q}={\bm k} - {\bm k}'\;;
\]
so that ${\bm W}_\perp={\bm Q}-{\bm q}$.
With this notation we re-express the vector arguments of the reduced
$g$ matrix,
\[
g_{Z}({\bm K}_{||},{\bm b}',{\bm b})\to
g_{Z}({\bm K+\bm P},{\bm \kappa}_{-},{\bm \kappa}_{+})\;,
\]
where
\begin{equation}
\label{kappa}
{\bm\kappa}_{\pm}=
\textstyle{\frac{1}{2}}
[{\bm K}-{\bm P}\pm \textstyle{\frac{1}{2}}({\bm q+\bm Q}) ]\;.
\end{equation}
With these considerations and using Eq. (\ref{TheResult}) for the 
two-body interaction, the \emph{unabridged}
\footnote{By \emph{unabridged} we mean
the nine-dimensional integration comprised of
six dimensions in momentum space ($d{\bm P}d{\bm Q}$) and
three in coordinate space ($d{\bm Z}$).
Spherical mass distribution reduces the dimension of non trivial
integrals to seven (6+1).  }
optical potential takes the form
\begin{equation}
\label{u}
U=U_0+U_1\;,
\end{equation}
with
\begin{subequations}
\begin{eqnarray}
\label{kmt}
U_{0}&=&\int \;d{\bm P}\;\hat\rho(\bm q;\bm P)\;g_\infty\;,\\
\label{unabridged}
U_{1} &=& {\;-\;}
\frac{1}{2\pi^2}
\int d{\bm Q}\;d{\bm P}\;
\hat\rho(\bm Q;\bm P)\; \times \nonumber \\
& & \int_{0}^{\infty}Z^3 dZ\;
\Phi_1(Z|\bm Q-\bm q|)\;
\frac{\partial g_Z}{\partial Z}
\;.
\end{eqnarray}
\end{subequations}
The first term, $U_{0}$, depends exclusively on the medium-free 
reduced matrix, whereas the second depends on the gradient of $g$.

\section{Nuclear matter model}
In the preceding analysis we have made no mention to a specific
approach to model the $\hat T$ matrix. In this regard, the framework
is general enough to include various strategies to describe an 
effective two-body interaction in the realm of a finite nucleus.
However, if the reduced matrix $g$ is taken \cite{Are95} as the 
antisymmetrized Brueckner-Bethe-Goldstone reaction matrix of starting 
energy $E$, 
\[
g(E)=v+v\;\frac{\hat Q}{E+i\eta-\hat h_1-\hat h_2}\;g(E)\;,
\]
then the reduced matrix at infinity, $g_\infty$, corresponds to the free 
scattering matrix $t(E)$. 
In the above equation $\hat h_1$ and $\hat h_2$ correspond to
quasi-particle energies and $\hat Q$, the Pauli blocking operator.
Therefore, the first term of the optical potential in Eq. (\ref{u}) becomes
\[
U_0(E)=\int d{\bm P}\;\hat\rho(\bm q;\bm P)\;t(E)\;,
\]
the lowest-order free $t$ matrix full-folding optical potential in
the Watson and Kerman-McManus-Thaler approach \cite{Ker59,Are89}.
Actual calculations of this contribution were realized in the early
nineties \cite{Are90a,Els90,Cre90}.

With regard to $U_{1}$ we note that the $g$ matrix in 
Brueckner-Bethe-Goldstone approach is a functional of the density, 
$g=g[\rho]$. 
If the reduced matrix is evaluated at a density $\rho$ specified by
the local coordinate ${\bm Z}$, then
\begin{equation}
\label{dgdz}
\frac{\partial g_Z}{\partial Z}=
\left (\frac{\partial\rho}{\partial Z}\right)\times
\frac{\partial g}{\partial\rho}\;.
\end{equation}
Considering that $\rho'(Z)\equiv \partial\rho/\partial Z$ peaks 
in the nuclear surface, the intrinsic medium-dependent contributions to
the optical potential become accentuated in that region.
The strength of such contributions will depend on $\partial g(E)/\partial\rho$,
an energy-dependent operator in spin-isospin space.

\begin{figure}%[ht]
\includegraphics[scale=0.45,angle=00] {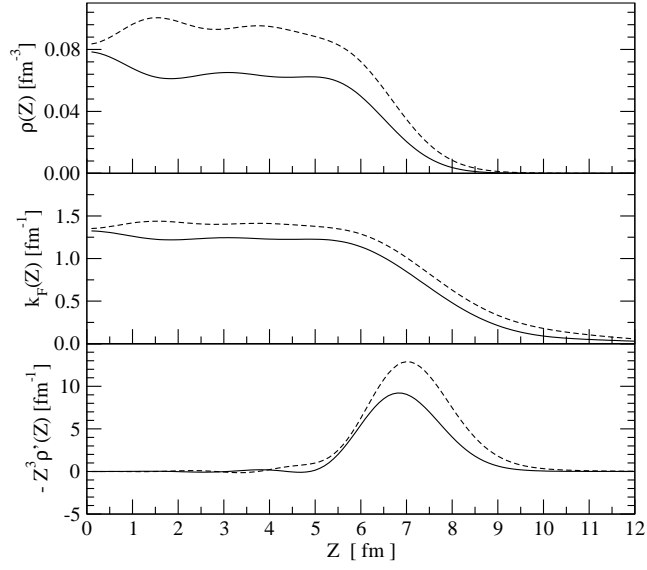}
\medskip
\caption{{\protect\small
\label{DensityProfile}
  Local density $\rho$ (upper frame), local momentum $\hat k_F$
 (middle frame) and $Z^3$ times the negative gradient of the density
(lower frame) as functions of the local coordinate $Z$. 
 The solid and dashed curves correspond to protons and neutrons in
 $^{208}$Pb, respectively.}
        }
\end{figure}

To focus these ideas, in Fig. (\ref{DensityProfile}) we characterize 
the proton and neutron densities in $^{208}$Pb \cite{Ber91}, where in the upper 
frame we plot the proton and neutron density, in the middle frame the 
local Fermi momentum given by
\[
k_F = (3\pi^2\rho)^{1/3}\;,
\]
and in the lower frame the negative gradient of the density. 
We have multiplied this function by $Z^3$ to account for its 
actual weight in the radial integration.

What becomes clear from this figure is that medium effects stemming 
from $\partial g/\partial\rho$ become dominant (if non zero) in 
the region between 5.5 fm and 9 fm. 
In this range the local Fermi momentum $k_F$ varies between 0.3 fm$^{-1}$ 
and 1.2 fm$^{-1}$, suggesting the densities at which the $g$ matrix needs 
to evaluated.

\subsection{Contact with the ABL approach}

The ABL approach to the optical potential constitutes an extension of
the early full-folding approach based on the free $t$ matrix \cite{Are95}. 
This extension makes use of an infinite nuclear-matter model to represent 
the \emph{NN} effective interaction between the projectile and the target
nucleons. Upon the use of a simplifying assumption regarding
its relative momenta dependence, and resorting to the Slater 
approximation of the mixed density, the optical potential takes 
the form of a folding of the diagonal (local) density with a nonlocal 
Fermi-averaged $g$ matrix. 
We verify this result as a limit case of Eq.  (\ref{unabridged})
for the unabridged optical potential.

If we assume that the $g$ matrix in Eq.  (\ref{unabridged}) exhibits 
a weak dependence on the transferred momentum ${\bm Q}$, 
then the relative momenta in $g$ can evaluated at ${\bm Q}={\bm q}$, 
consistent with the peak of $\Phi_1$ at the origin.
With these considerations we set 
\begin{equation}
\label{kappa0}
{\bm\kappa}_{\pm}\to
{\bm\kappa}_{\pm}^{(0)}\equiv
\textstyle{\frac{1}{2}}
({\bm K}-{\bm P}\pm {\bm q}) )\;,
\end{equation}
and symbolize
\[
g_Z(E)\to g_Z^{(0)}(E)
\]
Additionally, if we use (\ref{rhoqp}) for the Slater approximation to 
$\hat\rho(\bm Q;\bm P)$, then
\begin{eqnarray}
U_{1}
&=&
{\;-\;}
\frac{2}{\pi}
\int_{0}^{\infty}Z^3 dZ\;
\int_{0}^{\infty}{Z'}^2 dZ'\;\rho(Z')\;
\times
\nonumber \\
& &
\int d{\bm Q}\;
\Phi_1(Z|\bm Q-\bm q|)\;
j_0(QZ')\; 
\times
\nonumber \\
& &
\int d{\bm P}\;
\frac{\partial g_Z^{(0)}(E)}{\partial Z}
S_F(P;Z')\;.
\end{eqnarray}
Here the rightmost integral does not depend on ${\bm Q}$, therefore
the $d{\bm Q}$ integration involving $\Phi_1\;j_0(QZ')$ can be performed
separately.
Using Eq. (\ref{Wintegral}) in Appendix \ref{Omega}
and reordering the integrals we obtain
\begin{eqnarray}
\label{u1}
U_{1}
&=&
{\;-\;}
4\pi
\int_{0}^{\infty}{Z'}^2 dZ'\;j_{0}(qZ')\,\rho(Z')\;
\int d{\bm P}\;S_F(P;Z')\;
\times
\nonumber \\
& &
\int_{0}^{\infty}dZ\;
\Theta(Z-Z')\;
\frac{\partial g_Z^{(0)}(E)}{\partial Z}\;.
\end{eqnarray}
The integration over $Z$ is immediate. 
If we identify $g_{\infty}^{(0)}(E)=t(E)$,
the free $t$ matrix, then 
\begin{eqnarray}
U_{1}
&=&
{\;-\;}
4\pi
\int_{0}^{\infty}{Z'}^2 dZ'\;j_{0}(qZ')\,\rho(Z')\;\times
\nonumber \\
& &
\int d{\bm P}\;S_F(P;Z')\;
[t(E)-g_{Z'}^{(0)}(E)]\;.
\end{eqnarray}
The integral involving the free $t$ matrix leads to $U_{0}(E)$,
whereas the one involving $g_{Z'}^{(0)}(E)$ leads to the ABL 
\emph{in-medium} folding potential $U_{ABL}(E)$,
$ U_{1}\to -U_{0}+ U_{ABL}$, with
\begin{equation}
\label{abl}
U_{ABL}(E)=4\pi
\int_{0}^{\infty}{Z'}^2 dZ'\;j_{0}(qZ')\,\rho(Z')\;g_{Z'}^{(0)}(E)\;.
\end{equation}
When this term is substituted in Eq. (\ref{u}) we obtain
$U(E) \to U_{ABL}(E)$, the expected limit.
As seen, this result illustrates the fact that $U_{ABL}$ can be
rigorously derived from Eq. (\ref{unabridged}) using a few simplifying
assumptions, i.e. Slater approximation of the mixed density and
a weak dependence of the $g$ matrix in the ${\bm Q}$ momentum.

\subsection{Contrast with the $\rho(\partial/\partial\rho)$ rearrangement term}
During the course of this work it was noticed by some colleagues
certain resemblance between Eq. (\ref{unabridged}) for the unabridged 
optical potential, and medium corrections in the form of a 
$\rho(\partial/\partial\rho)$ term proposed by Cheon and 
collaborators \cite{Che84}.
It becomes appropriate, therefore, to point out the differences in context and
form of this apparent resemblance.

The $\rho(\partial/\partial\rho)$ term of Cheon \emph{et al.}
emerges after a perturbative treatment of the transition density 
for inelastic scattering.
In this case the optical potential for elastic scattering is given 
schematically by $U_{opt}=G(\rho)\rho$, 
with $G(\rho)$ the Brueckner $G$ matrix and $\rho$ the ground 
state mixed density. 
The transition potential for inelastic scattering, $U_{tr}$, is 
expressed as $U_{tr}=(\partial U_{opt}/\partial\rho)\rho_{tr}$, 
with $\rho_{tr}$ the transition density.
Combining these equations it is shown that the transition potential can
be expressed as
\[
U_{tr}=
\left [
G(\rho)+\rho\frac{\partial G(\rho)}{\partial\rho}
\right ]
\rho_{tr}\;.
\]
As observed, the optical potential $U_{tr}$ results the sum of
the elastic term with a corrective term of the form
$\rho(\partial/\partial\rho)$. This correction accounts for the
rearrangement of the target nucleons in an inelastic process.
Additionally, the $\rho(\partial/\partial\rho)$ form of the
correction implies relatively uniform contributions in the 
nuclear interior, in contrast with the 
$\rho\,'(\partial/\partial\rho)$ term in Eq. (\ref{unabridged})
for \emph{elastic scattering}, where the intrinsic medium 
effects manifest dominantly in the nuclear surface.
The unabridged optical potential discussed here represents 
elastic processes and its extension to inelastic
scattering would require further analysis.

\subsection{Medium sensitivity}
The actual evaluation of the unabridged optical potential
[c.f. Eq. (\ref{unabridged})] constitutes a very challenging task
beyond the scope of this work.
Indeed, each matrix element requires the realization of a 7-dimensional
integration, three more dimensions than current calculations in the
ABL approach.
However, it is possible to assess the relative importance of
selected terms in the medium-dependent $U_1$ contribution.
In order to isolate the role of medium effects, let us define
the amplitude
\begin{equation}
\Gamma(Z';E,Z)
\equiv
\int d{\bm P}\;S_F(P;Z')\;
\frac{\partial g_Z^{(0)}(E)}{\partial Z}\;.
\end{equation}
In the context of Eq. (\ref{u1}) this amplitude accounts for 
the Fermi average of the gradient of the effective interaction in the limit 
${\bm\kappa}_{\pm}\to {\bm\kappa}_{\pm}^{(0)}$. 
Thus, it is reasonable to expect that this quantity accounts for
the leading contributions stemming from the $d{\bm Q}$ integral in
Eq. (\ref{unabridged}). 

\begin{figure}[ht]
\includegraphics[scale=0.45,angle=00] {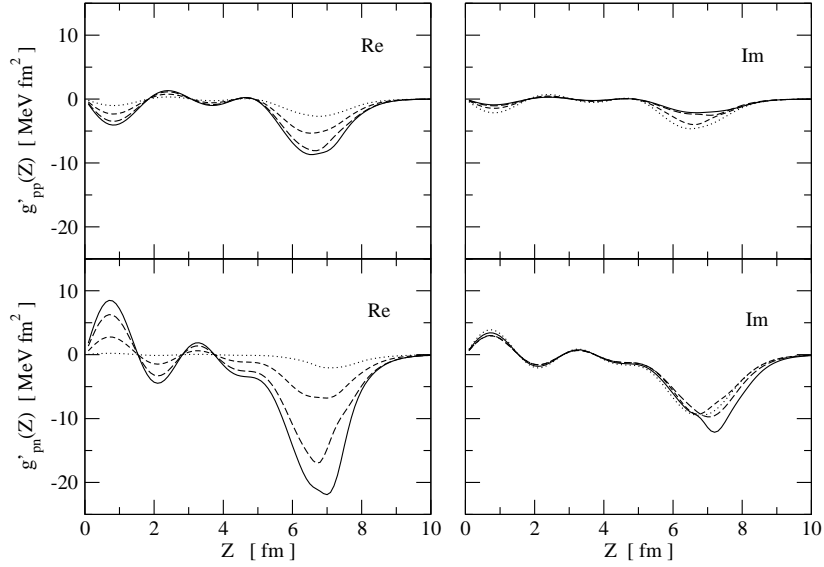}
\medskip
\caption{{\protect\small
\label{gradient}
  Fermi integral of the gradient of the $g$ matrix ($\Gamma_{NN}$) 
  as function of the radial distance $Z$ in $^{208}$Pb. 
  The \emph{pp} (\emph{pn}) channel is shown in the upper (lower) frame,
  whereas the real (imaginary) components are plotted in the left
 (right) panels.
  The solid, long-dashed, short-dashed and dotted curves represent results
  for $\Gamma_{NN}$ at nucleon energies of 65, 100, 200 and 300 MeV, 
  respectively.
 }
        }
\end{figure}

In addition to $Z'$, $Z$ and $E$, the $\Gamma$ amplitude depends on 
the momenta ${\bm k}$ and ${\bm k}'$, and spin-isospin degrees of freedom.
To examine its radial behavior, we have evaluated
the diagonal ($Z=Z'$) elements which are plotted in 
Fig. (\ref{gradient}) for the central \emph{pp} (upper frames) and 
\emph{pn} (lower frames) channels.
The real and imaginary components are shown in the left and right 
panels, respectively.
These amplitudes are evaluated on-shell ($E=k^2/2m={k'}^2/2m$),
at forward angles, for nucleon energies of 
65 MeV (solid curves), 
100 MeV (long-dashed curves), 
200 MeV (short-dashed curves) and 
300 MeV (dotted curves).
The density considered in this analysis is that of $^{208}$Pb, 
as in Fig. (\ref{DensityProfile}).

By simple inspection we observe that medium effects
accounted for by $U_{1}$ add attraction and absorption to the 
medium-free $U_{0}$ contribution.
All of them are localized in the nuclear surface,
becoming weaker in the interior. 
Additionally, the various components of $\Gamma$ exhibit distinctive 
features regarding their energy and medium sensitivity.
The energy dependence is identified by the separation of
among all four curves, being this most pronounced in the case of 
$\textrm{Re}\,\Gamma_{pn}(E)$ and followed by 
$\textrm{Re}\,\Gamma_{pp}(E)$.
The weakest energy dependence occurs for the
absorptive component in the \emph{pn} channel.

Regarding the intrinsic medium effects, the strongest dependence 
occurs for the real \emph{pn} amplitude at 65 MeV.
In contrast, the weakest dependence occurs for the absorptive \emph{pp} 
(upper-right frame) amplitude. 
Additionally, with the exemption of $\textrm{Im}\,\Gamma_{pn}(E)$,
all the other amplitudes exhibit decreasing strength with increasing energy.

%%%%%%%%%%%%%%%%%%%%%%%%%%%%%%%%%%%%%%%%%%%%%%%%%%%%%%%%%%%%%%%%%%%%%
\section{Summary and conclusions}
We have examined the role of medium effects in the optical
model potential for hadron-nucleus scattering. 
The analysis is based on a close scrutiny of the structure of the
two-body effective interaction with spherical symmetry in its 
local coordinate ${\bm Z}$, leading to its asymptotic separation.
As a result, we demonstrate that the unabridged optical potential
can be separated into two very distinctive contributions.
One of them is a momentum-conserving and medium-independent term,
while the other is functionally proportional to the radial 
derivative of the reduced matrix element.
If the Brueckner-Hartree-Fock $g$ matrix is used to model the
\emph{NN} effective interaction, then the medium-independent 
term of the optical potential corresponds to the well known
Watson-KMT lowest order full-folding optical model potential,
while the medium-dependent term depends exclusively on the 
gradient of the reduced $g$ matrix.
The modulation by the radial derivatives of the density enhances that
effect in the nuclear surface and suppresses it in the saturated volume.

The assessment of the intrinsic medium effects by means of the
$\Gamma(Z';E,Z)$ amplitude points to stronger medium sensitivity
in the real part of the \emph{pn} amplitude at 65 MeV, in contrast 
with the absorptive \emph{pp} component. 
The energy-dependence of these effects are relatively weak in
the absorptive \emph{pp} and \emph{pn} channels.
The introduction of the $\Gamma$ amplitude may prove to be very
useful for the assessment of intrinsic medium effects, such as
those associated with the inclusion of non spherical components 
of the Pauli blocking studied in Ref. \cite{Ste05}. 

Although we have not provided a full realization of the unabridged
optical potential, we have been able to extract its exact functional 
dependence in terms of the nuclear density when the system is 
spherically symmetric. 
This feature may become useful in the context of semi-phenomenological 
approaches, where the free-space contribution may well be represented 
in terms of impulse-approximation-like potentials, while the medium 
dependent term can be modeled as function of 
$\rho'\partial/\partial\rho$ couplings. 
At a more fundamental level, it would be interesting to identify
missing features in the ABL approach relative to the unabridged potential. 
As inferred from Fig. (\ref{gradient}), intrinsic medium effects become
most pronounced --in the nuclear surface-- at lower energies.
This surface-sensitive phenomenon may be of particular importance
in the study of rare isotope beams, where highly unstable nuclei
are collided against hydrogen targets. 
In this case, the traditional intermediate energy regime is reached 
with 30A-100A MeV beams.  

\appendix
\section{Representation of quantum two-body operators}
\label{TwoBody}
Let $A$ a two-body operator and denote its coordinate representation by
$
\langle {\bm r}' {\bm s}'\mid A\mid{\bm r}\; {\bm s}\rangle =
A({\bm r}' {\bm s}';{\bm r}\; {\bm s})\;,
$
where ${\bm r},{\bm s}$ (${\bm r}',{\bm s}'$) represent the prior
(post) coordinates of each particle.
An alternative set of coordinates is summarized by Eqs. (\ref{rset1})
and illustrated  in Fig. (\ref{fourv}b).
With this transformation we can express 
$
A({\bm r}'{\bm s}';{\bm r}{\bm s})=A_{\bm R'\bm R}({\bm x}',{\bm x})\;.
$
If we denote
$\tilde A\equiv\langle {\bm k}' {\bm p}' \mid A\mid{\bm k}\; {\bm p}\rangle$,
then 
\begin{eqnarray}
\tilde A &=&
\frac{1}{(2\pi)^6}
\int d{\bm r}'d{\bm s}'d{\bm r}\;d{\bm s}\;
\times\nonumber\\
&&
e^{-i(
{\bm k}'\cdot{\bm r}'+
{\bm p}'\cdot{\bm s}'-
{\bm k}\cdot{\bm r}-
{\bm p}\cdot{\bm s})}
A({\bm r}'{\bm s}';{\bm r}\; {\bm s})\;.
\nonumber
\end{eqnarray}
In terms of the coordinate set defined in Eqs. (\ref{rset1}) we re-express
the above integral in the form
\begin{eqnarray}
\tilde A&=&
\frac{1}{(2\pi)^6}
\int\; d{\bm R}'\;d{\bm R}\;d{\bm x}'\;d{\bm x}\;
\times\nonumber\\
& &
e^{-i(
{\bm W}'\cdot{\bm R}'+
{\bm b}'\cdot{\bm x}'-
{\bm W}\cdot{\bm R}-
{\bm b}\cdot{\bm x})}
A_{\bm R'\bm R}({\bm x}',{\bm x})\;,
\nonumber
\end{eqnarray}
where we denote ${\bm W} = {\bm k} +{\bm p}$, the momentum of
the pair prior interaction and ${\bm b} = ({\bm k} -{\bm p} )/2$, the
relative momentum. Here again prime marks denote post interaction
as shown in Fig. (\ref{fourm}).
To the above expression we apply the Wigner transform in the prior
and post center of mass coordinates ${\bm R}$ and ${\bm R'}$.
Defining ${\bm Y}= \bm R'-\bm R$, and ${\bm Z}=(\bm R'+\bm R)/2$ we obtain
\begin{equation}
\label{wigner2}
\tilde A=
\int\;
\frac{d{\bm Z}}{(2\pi)^3}
\;e^{i{\bm Z}\cdot({\bm W}'-\bm W)}\;
g_{\bm Z}[\textstyle{\frac12}(\bm W'+\bm W); {\bm b}',{\bm b}]\;,
\end{equation}
where $g_{\bm Z}$ represents a reduced interaction $g$ at the 
coordinate ${\bm Z}$ and is given by
\begin{eqnarray}
\label{located}
g_{\bm Z}({\bm K}_{||};{\bm b}',{\bm b}) 
&=&
\int \frac{d{\bm Y}}{(2\pi)^3}
e^{-i{\bm Y}\cdot{\bm K}_{||}} \int
d{\bm x}'d{\bm x}\;
e^{-i({\bm b}'\cdot{\bm x}'-{\bm b}\cdot{\bm x})}
\times \nonumber \\
&&
A_{\bm Z +\frac12\bm Y,\bm Z -\frac12\bm Y}({\bm x}',{\bm x})\;.
\nonumber
\end{eqnarray}

\section{The ground-state mixed density}
\label{Mixed}

Let us denote the one-body density matrix in momentum space by
$\tilde\rho({\bm p}',{\bm p})$. In coordinate space this matrix
is represented by $\rho({\bm r}',{\bm r})$, where
\begin{equation}
\tilde\rho({\bm p}',{\bm p})=
\frac{1}{(2\pi)^3}\int d{\bm r}'d{\bm r}\;
e^{-i{\bm p}'\cdot{\bm r}'}
\rho({\bm r}',{\bm r})\;e^{i{\bm p}\cdot{\bm r}}\;.
\nonumber
\end{equation}
It is customary to introduce the center-of-mass ${\bm Z}=({\bm r}'+{\bm r})/2$, 
and relative ${\bm x}={\bm r}'-{\bm r}$, coordinates. 
Similarly, we define the mean ${\bm P}=({\bm p}'+{\bm p})/2$, 
and transferred ${\bm Q}={\bm p}'-{\bm p}$, momenta.
With the use of these transformations we denote
$ \tilde\rho({\bm p}',{\bm p})\equiv \tilde\rho({\bm Q};{\bm P})\;;
\rho({\bm r}',{\bm r})\equiv \rho({\bm Z};{\bm x})\;.  $ 
Therefore,
\begin{eqnarray}
\tilde\rho({\bm Q};{\bm P})&=&
\frac{1}{(2\pi)^3}\int d{\bm Z}\;d{\bm r}\;
e^{-i({\bm P}\cdot{\bm x}+{\bm Q}\cdot{\bm Z})}\rho({\bm Z};{\bm x}) 
\;;\nonumber\\
&\equiv&
\frac{1}{(2\pi)^3}\int d{\bm Z}\; e^{-i{\bm Q}\cdot{\bm Z}}\;\rho({\bm Z})\;
G({\bm Z};{\bm P})\;,
\label{rhoqp}
\end{eqnarray}
where we have defined
\[
\rho({\bm Z})\;G({\bm Z};{\bm P})=
\int d{\bm x}\;e^{-i{\bm P}\cdot{\bm x}}\;\rho({\bm Z};{\bm x})\;.
\]

In the Slater approximation for the mixed density its coordinate 
representation takes the form
\begin{equation}
\rho({\bm Z};{\bm x})=\rho({\bm Z})\;F(Z;x)\;,
\end{equation}
with $F(Z;x)=3\;j_1(\hat k_Zr)/(\hat k_Zr)$. 
Here $\hat k_Z$, the local Fermi momentum, depends on the local 
density $\rho(Z)$ through
\[
\hat k = (3\pi^2\rho)^{1/3}\;.
\]
Within this approximation, and assuming spherical symmetry in the local
density, the function $G({\bm Z};{\bm P})$ can be calculated directly.
When used to evaluate Eq. (\ref{rhoqp}) for $\tilde\rho({\bm Q};{\bm P})$
we obtain
\begin{equation}
\tilde\rho({\bm Q};{\bm P})=4\pi\int_{0}^{\infty}Z^2dZ\;j_0(QZ)\;\rho(Z)\;
S_F(P;Z)\;,
\end{equation}
where
\begin{equation}
S_F(P;Z)=\frac{1}{\frac{4}{3}\pi \hat k_Z^3}\;\Theta(\hat k_Z - P)\;.
\end{equation}

\section{Evaluation of $\int d{\bm Q} \; j_{0}\Phi_1$}
\label{Omega}
We evaluate the volume integral $\Omega_q(Z',Z)$ defined by
\[
\Omega_q(Z',Z)\equiv 
\int d{\bm Q} \; j_{0}(QZ')\;
\left [
\frac{j_{1}(Z|\bm Q-\bm q|)}{Z|\bm Q-\bm q|}
\right ]\,
\]
where we express $|\bm Q-\bm q|=\sqrt{Q^2+q^2-2Qqu}$, 
with $u=\hat Q\cdot \hat q$.
The solid-angle integration is straightforward, leading for $\Omega$
\begin{eqnarray}
\Omega_q(Z',Z)&=&
\frac{4\pi}{Z^2Z'}\int_{0}^{\,\infty}\frac{Q\,\sin QZ'\,dQ}{q^2-Q^2}\,
\times  \nonumber \\
&& \left [
\frac{\cos(QZ)\sin(qZ)}{qZ}-\frac{\sin(QZ)\cos(qZ)}{QZ}
\right ] \;. \nonumber
\end{eqnarray}
This integral can be evaluated analytically and the expression 
depends on the location of $Z$ relative to $Z'$. 
The two cases of interest are $Z>Z'$, and $Z<Z'$, where
we obtain
\begin{equation}
\Omega_q(Z',Z)=\left\{
\begin{array}{lr}
0 & \textrm{if}\quad Z<Z';\\
\left (\frac{2\pi^2}{Z^3}\right )\,j_{0}(qZ')& \textrm{if}\quad Z>Z';
\end{array}
\right .
\end{equation}
When $Z=Z'$ the result is finite but discontinuous.
Due to its involved form we have preferred to omit it
for clarity, keeping in mind that at $Z=Z'$ its 
contribution as integrand vanishes. 
Thus, without loss of generality we can express
\begin{equation}
\label{Wintegral}
\Omega_q(Z',Z)=\frac{2\pi^2}{Z^3}\,j_{0}(qZ')\,\times
\Theta(Z-Z')\;,
\end{equation}
with $\Theta(x)$ the Heaviside step function.

%%%%%%%%%%%%%%%%%%%%%%%%%%%%%%%%%%%%%%%%%%%%%%%%%%%%%%%%%%%%%%%%%%%%%

\begin{acknowledgments}
{\small H.F.A.} acknowledges partial support provided by {\small FONDECYT} 
under grant 1040938.
\end{acknowledgments}

\newpage

\end{document}